\documentclass[prl,twocolumn,showpacs,preprintnumbers,amsmath,amssymb]{revtex4} 
\usepackage{amsmath}
\usepackage{graphics}
\usepackage{graphicx}
\usepackage{dcolumn}
\usepackage{bm}

\begin{document}

%

\title{Disorder-Induced Vibrational Localization}

\author{J.~J.~Ludlam, S.~N.~Taraskin, S.~R.~Elliott }
\affiliation{ Department of Chemistry, University of Cambridge,
	      Lensfield Road, Cambridge, CB2 1EW, UK }

\date{\today}

\begin{abstract}

The vibrational equivalent of the Anderson tight-binding
Hamiltonian has been studied, with particular focus on the 
properties of the eigenstates at the transition from 
extended to localized states. The critical energy has been
found approximately for several degrees of force-constant disorder using
system-size scaling of the multifractal spectra of the
eigenmodes, and the spectrum at which there is no system-size 
dependence has been obtained. This is shown to be in good
agreement with the critical spectrum for the electronic
problem, which has been derived both numerically and by 
analytic means. Universality of the critical states
is therefore suggested also to hold for the vibrational problem.

\end{abstract}

\pacs{63.50.+x, 63.20.Pw}

\maketitle

The Anderson electron localization problem \cite{Anderson_58} is one that has attracted
much attention over the last 40 years. The fact that the problem
can be stated so simply, and yet have startlingly complex 
consequences, has made it a challenging topic to work on \cite{Kramer_93}. Indeed,
only recently has it been possible to verify numerically
many of the theoretical results on powerful supercomputers \cite{Schreiber_96:Book}. However, the 
closely related vibrational problem has not been explored to the same degree, 
despite being similar enough to use the same techniques 
yet different enough to produce new and interesting results.

The phenomenon of localization is a second-order phase transition 
between eigenstates that are spatially localized and those that are 
delocalized, or extended \cite{Janssen_98:review}. In the thermodynamic limit, extended 
eigenmodes would cover the whole of space whereas localized 
eigenstates are those which only involve a local subset of the 
system within a typical localization length. In the crystalline 
case for both the electronic and vibrational problems, 
the eigenstates are simple Bloch states due to translational 
invarience, and are therefore extended.
For the electronic problem, disorder is generally 
introduced either in the on-site energy terms (diagonal disorder) 
or the interaction terms (off-diagonal disorder) \cite{Kramer_93}. 
In a 3d lattice with weak diagonal disorder, 
there are two critical energies, at the top and bottom of the 
band, at which the Localization-Delocalization (LD) transition 
occurs. As the degree of disorder is increased, these two critical 
energies approach, and finally meet. At this point, all the 
eigenmodes are localized and the system becomes an electrical 
insulator. This transition is termed the Metal-Insulator 
Transition (MIT) \cite{MacKinnon_81}. Off-diagonal disorder produces fundamentally
different behaviour: at no level of disorder are all the eigenmodes
localized, and hence there is no MIT \cite{Cain_99}.

Our approach to the problem of vibrational localization has been numerical, 
applying high-perfomance computers to the task of obtaining the eigenmodes. 
The Anderson
electron Hamiltonian can be expressed in a site basis, giving a
sparse matrix representation of the problem, for which the eigenvectors
can then be 
found by using standard Lanczos methods. 
Modern computers can solve such eigenproblems for many millions of 
atomic sites. A bigger problem is how to recognise quantitatively
the difference between localized and extended states.

There are several methods for distinguishing extended from
localized states, e.g. by looking at the properties of the Hamiltonian, 
such as the transfer matrix method 
\cite{MacKinnon_81,Pichard_81}, 
observing differences in the level-spacing statistics \cite{Carpena_99},
the Thouless criterion \cite{Thouless_72}, or by looking at
the eigenstates themselves. The latter is not trivial though, since
as the critical energy is approached from the localized regime, 
the localization length diverges. Thus, for a finite system size, 
the eigenmodes quickly become extended over a larger range than 
the system size and it becomes difficult to assess whether a state 
is truly localized or extended. These states are known as 
prelocalized states \cite{Mirlin_00:review}, and to characterize
these as localized or extended, we can use multifractal analysis (MFA)
\cite{Janssen_94}. 

It has been suggested that the eigenmode at exactly the LD critical 
energy will show multifractal characteristics 
\cite{Schreiber_91}. The standard way of characterising the 
multifractality is the singularity spectrum, which has been shown 
for electrons \cite{Grussbach_95} to be universal for an isotropic
system (see Ref.\cite{Milde_97} for treatment of an anisotropic system) 
and independent 
of the probability distribution of the disorder. 
The analytic predictions for the singularity spectrum \cite{Wegner_89}, 
based on the $d=2+\epsilon$ expansion of the non-linear $\sigma$ model, 
are in good agreement with numerics \cite{Grussbach_95}.

The aim of this paper is two-fold. Firstly, we use MFA in order to 
identify the threshold energy of the LD transition for different
degrees of force-constant disorder and thus obtain the ``phase diagram'' 
 in the frequency-disorder
plane for vibrational excitations in disordered models.
Secondly, we demonstrate the universal features of the multifractal
critical states at the LD transitition for the vibrational problem.

We can use the idea of critical multifractality to determine whether  
the states are extended or localized by looking at how the 
singularity spectrum, characterising the eigenmode, changes with 
simulation-system size. For a true multifractal state, assuming
that finite-size effects are small, the 
singularity spectrum will not depend on the simulation box size, 
whereas the spectra for states on either side of the LD 
transition will vary. Hence, by calculating the singularity 
spectrum for different system sizes, we can locate the critical 
energy \cite{Grussbach_95}. 

The harmonic vibrational problem that is addressed in this Letter can be 
formulated in a very similar way to the Anderson electron problem \cite{Elliott_74}.
For vibrations, the equivalent to the electronic Hamiltonian 
is the symmetric dynamical operator: 
\begin{equation}
\label{dynmat}
{\hat D}=\sum\limits_{(i\alpha)(j\beta)}
D_{(i\alpha)(j\beta)}
(\left| i,\alpha\right\rangle - \left|j,\alpha\right\rangle)
(\left\langle i,\beta\right| - \left\langle j,\beta \right|)~,
\end{equation}
with $\left| i,\alpha \right\rangle$ being the site basis 
describing the displacement of atom $i$ ($i=1\ldots N$) along the Cartesian 
direction $\alpha$ ($\alpha=1\ldots d$, with $d$ the 
dimensionality). The matrix elements
$ D_{(i\alpha)(j\beta)} 
= (\kappa_{ij}/2) \left( 
{\hat r}_{ij} 
\right)_\alpha  
 \left({\hat r}_{ij}
\right)_\beta $ 
are defined in terms of force constants $\kappa_{ij}$, and unit vectors, 
${\hat r}_{ij}$, connecting the atoms $i$ and $j$ (for simplicity, 
all masses are taken to be equal, $m_i=1$).  The dynamical matrix 
consists of $ d\times d$ blocks with strong lattice symmetry-dictated 
correlations inside the blocks. Additionally, 
all the elements of the 
on-diagonal blocks are the sums (with opposite sign) of the 
similar elements of off-diagonal blocks, 
reflecting the sum-rule correlations in the dynamical matrix. Therefore, 
in the case of nearest-neighbor interactions considered below, the
number of correlations between the elements in the dynamical matrix
is comparable with the number of independent random variables.

There are three main differences between the Anderson electron and 
the vibrational problems. Firstly, when there are no negative 
values of $\kappa_{ij}$, the system is mechanically stable, so there
are no negative eigenvalues, unlike the electron case. 
Secondly, the basic Anderson formulation 
gives a symmetric band structure. The vibrational case is  asymmetric 
for the third reason: there are $d$ zero-frequency modes that cannot 
be localized since they correspond to bulk translational 
displacements of the system (Goldstone modes). Since the lower bound
of the spectrum is therefore constrained to be extended in character,
we expect in a single-band model that there will be only one LD 
transition near the high-frequency band edge.
 
There are two major classes of model which can be used in studying 
localization: structures based on an underlying crystalline lattice 
with introduced disorder, and structures which have been 
created in an effort to recreate the distribution of atomic positions 
and bond angles found in real amorphous materials. 
For our study, we have chosen to analyse lattice models from the first class, 
with an underlying f.c.c. geometry and with the $\kappa_{ij}$ in 
Eq.~(\ref{dynmat})
taken from a probability distribution $\rho(\kappa)$. 
This is one of the 
simplest models and can be easily compared with the established 
results for the electron-localization problem for similar models. 
The distribution $\rho(\kappa)$ has been chosen to be a uniform (box) distribution, centered at 
$\kappa_0=1$ with a full width $\Delta < 2\kappa_0$ in order to give 
both a simple random 
distribution and one where there are no negative force constants.
Our models are face-centered cubic and range in size from $L=16$ with 4096 atoms
up to $L=48$ 
with $N= 110592$ atoms.


A multifractal is a generalization of a standard geometric fractal
for the case
when a single fractal dimension cannot characterize the
system \cite{Peitgen_92:book}. For each point in our measure, we can define
a value $\alpha(r)$ that describes the scaling of the
measure with $L$ around that point. If we now take
the set of all points with a specific $\alpha$, that
itself is a fractal, with dimensionality $f(\alpha)$.
The curve $f(\alpha)$ is known as the multifractal spectrum, or singularity spectrum, and can be used to characterize eigenmodes as localized or extended, as
shown below.
To calculate the scaling exponents, we define the measure
\(P_i(L_{b})=\sum _{j \in \text{Box}_i(L_b)} \left| \mathbf{u_j}\right| ^{2} \)
as the sum of the squared displacements $\left| \mathbf{u_j}\right|^2$ of
all the atoms $j$ within the $i$th box of size \( L_{b} \le L \) for a 
particular eigenmode, and examine
how this measure scales with $L_b$, or equivalently, with $\lambda=L_b / L$. 
We split our system up into $N(\lambda)$ boxes which completely and exactly
cover the system, so that $N(\lambda)=\lambda^{-d}$. 
The standard normalization of the eigenmodes 
leads to a scaling law for the measure of
the form \( \left\langle P(L_{b})\right\rangle _{L}\propto \lambda ^{d} \), 
averaging over all boxes.

The assumption underlying multifractal analysis is that, for a finite
interval of \( \lambda  \), the $q$th moments of the \( P(L_{b}) \)
also scale with power laws:
$\left\langle P^{q}(L_{b})\right\rangle _{L}\propto \lambda ^{d+\tau (q)}$
where \( \tau (q) \) is independent of \( \lambda  \). The range
of \( \lambda  \) in our case has a lower bound at the interatomic
spacing, since we are dealing with a discrete rather than a continuous
system. The upper bound $L/2$ is dictated by finite-size effects.
In the thermodynamic limit, as \( L\rightarrow \infty  \) ($\lambda\rightarrow 0$), 
the states which satisfy the multifractal condition are only found exactly at the critical energy,
and thus the exponents are defined uniquely as

\begin{equation}
 \label{tau}
\tau (q)=\lim _{\lambda \rightarrow 0}\frac{\ln (\left\langle P^{q}(L_{b})\right\rangle _{L})}{\ln \lambda }-d.
\end{equation}

In practice, $\tau (q)$ is found by performing a linear regression of the
calculated exponents with \( \ln \lambda  \). From this we can obtain
the singularity spectrum, \( f(\alpha ) \), where \( \alpha  \) is
defined as $\alpha (q)=\textrm{d}\tau(q)/\textrm{d} q$
and \( f(\alpha ) \) is obtained from the Legendre transformation
of \( \tau (q) \), \( f(\alpha (q))=\alpha (q)q-\tau (q) \).
Calculation of the singularity spectrum using the Legendre transformation
suffers from numerical errors, so it is more convenient to calculate
\( f(\alpha ) \) as a function of \( P(L_{b}) \) explicitly \cite{Peitgen_92:book}:
\begin{equation}
\label{ParametricAlpha}
\alpha (q)=\frac{1}{\ln \lambda }\sum _{\textrm{boxes}}\frac{P_{i}^{q}(L_{b})}{Z(q,L_{b})}\ln P_{i}(L_{b}),
\end{equation}
\begin{equation}
\label{ParametricF}
f(q)=\frac{1}{\ln \lambda }\sum _{\textrm{boxes}}\frac{P^{q}_{i}(L_{b})}{Z(q,L_{b})}\ln P_{i}^{q}(L_{b}),
\end{equation}
where $Z(q,L_b)=\sum_{\textrm{boxes}} P^{q}_i(L_b) $.

Since we cannot take the limit $\lambda \rightarrow 0$ in Eqs.(\ref{ParametricAlpha})-(\ref{ParametricF}), the values of \( f(q) \) 
and \( \alpha (q) \) are calculated
by performing a linear regression of the respective sums with respect to
\( \ln \lambda \). The linearity of these graphs is a good check of the 
multifractal nature of the measure.


Care has to be taken over the box sizes used in the analysis.
For example, taking the box to include just one atomic site proved to 
skew the regression, as did taking the box size be that of the entire
system. The reason for the former is that the multifractality must break
down at some point, certainly for box sizes on the order of the atomic
spacing. Finite-size effects account for the discrepancy for the largest
box size.

The singularity spectra of the eigenmodes around the critical energy fluctuate
strongly, and so it becomes necessary to take an average. Ideally, we would
like to average over different realizations of disorder, but in practice this
is only realistic for the smaller size models. For larger models, we take the
computationally cheaper option of averaging consecutive eigenmodes, which can
be obtained simply in the Lanczos algorithm. In order to reduce errors, we have
used the gliding-box method, averaging over all possible origins when dividing
the system into boxes \cite{Peitgen_92:book}.

\begin{figure}
\centerline{\includegraphics[width=\columnwidth]
{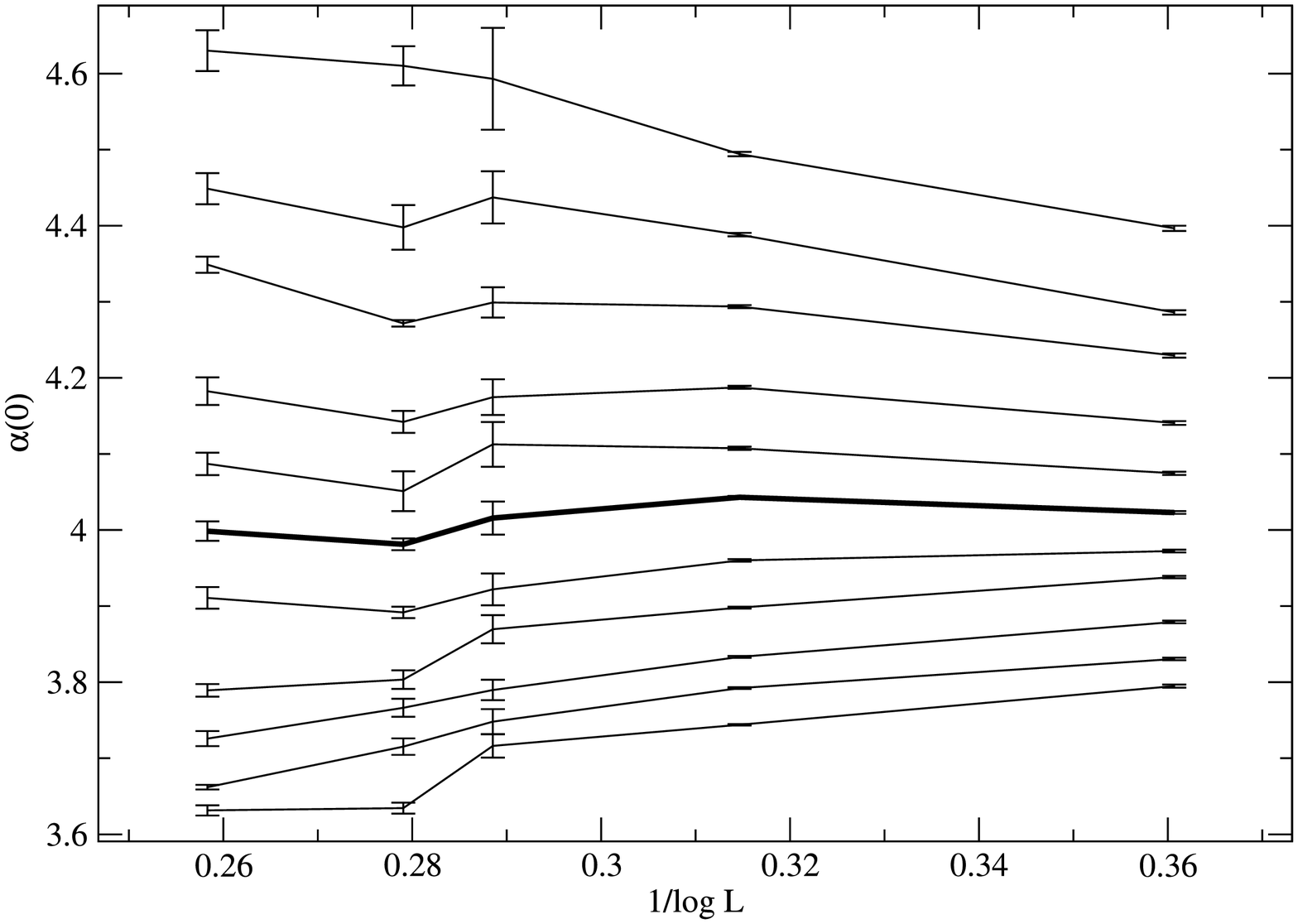}} 
\caption{\label{regressions}Estimation of the localization edge for $\Delta=1.5$. 
Each line is 
at a different frequency, from $\omega^2=9.3$ at the bottom to $9.5$ at the top in steps of $0.02$. The critical frequency is that at which this line has
zero gradient.
Notice the bold line shown, with approximately zero gradient, 
corresponds to $\omega^2=9.4$ and is at $\alpha(0)=4.0$}
\end{figure}

\begin{figure}
\centerline{\includegraphics[width=\columnwidth]
{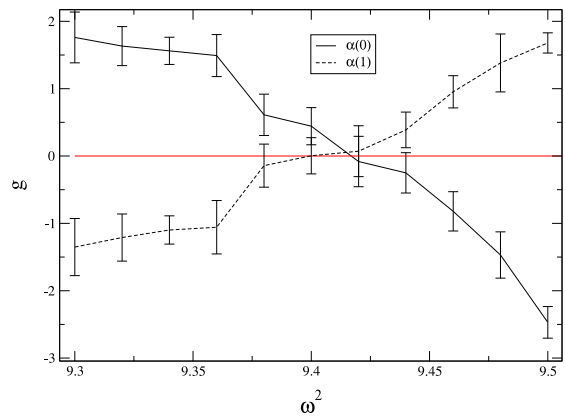}} 
\caption{\label{gradsvse}Plot of $g=\textrm{d}\alpha(q)/\textrm{d}(log{L})^{-1}$ for $q=0$ and $1$.
The squared critical frequency $\omega_*^2$ is given by the zero-crossing point of the graph.
In this case, $\omega_*^2$ is between 9.4 and 9.44. } 
\end{figure}

Once we have the spectra, we can find the frequency at which there is no change
with system size to locate the mobility edge. 
Empirically, it was noted \cite{Milde_97} that, for the Anderson
case, a plot
of \( \alpha(q) \) against \( (\log{L})^{-1} \) gave a good linear fit with
a different sign of gradient $g(\omega^2)=\textrm{d}\alpha(q;\omega^2)/\textrm{d}(log{L})^{-1}$ 
on either side of the critical frequency. The same
holds true for vibrational models, as clearly demonstrated in 
Fig. \ref{regressions}. We have therefore performed a 
linear regression on these curves, and the gradients of these lines have been
plotted at different energies to find the point where the singularity
spectrum is size independent, at which $g(\omega^2)$ crosses the abscissa. 
We can get additional information by looking at different values of $q$. 
In practice, since $\alpha(q)$ is strongly correlated for similar $q$, we
have looked at the representative values $q=0$ and $1$, 
for which the $g(\omega^2)$ have opposite
signs (see Fig. \ref{gradsvse}). 

\begin{figure}
\centerline{\includegraphics[width=\columnwidth]
{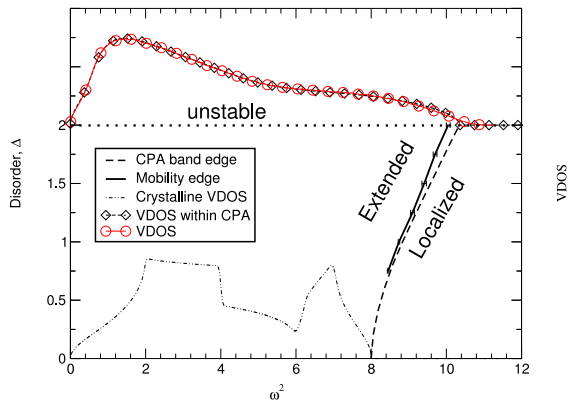}} 
\caption{\label{phasediagram} 
Phase diagram showing the boundary between 
extended and localized states. 
The VDOS for the crystal and the lattice with $\Delta=2.0$ 
have also been plotted to show the location of 
the mobility edge, $\omega_*^2$, within the band tail. 
The band edge calculated with CPA is also shown for reference.}
\end{figure}

\begin{figure}
\centerline{\includegraphics[width=\columnwidth]
{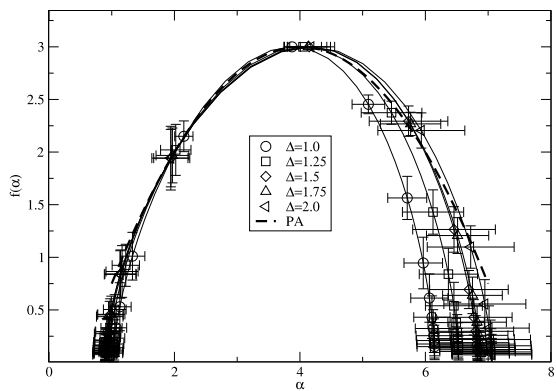}} 
\caption{\label{critspecs} Critical spectra for the force-constant disordered 
 models. 
The parabolic approximation (PA) to Wegner's result 
\cite{Wegner_89} is shown for comparison.} \end{figure}

Initially, the analysis was undertaken throughout the acoustic band. We did 
not expect
to find localization at the lower (zero-frequency) band edge \cite{Taraskin_01:PRL}, and indeed 
it was found that there was only one LD phase transition, located in the 
far high-energy band tail. The band edge calculated within the coherent potential
approximation (CPA) was found to be quite close to the true localization
threshold, as can be seen in Fig.\ref{phasediagram}, and therefore it can be 
used as a rough estimate for the frequency
of the actual LD transition .

Having found the mobility edge for several values of force-constant 
disorder $\Delta$, we
can plot these to produce a 'phase diagram' of the eigenmodes. This is shown
alongside the VDOS and the CPA band edge in Fig.\ref{phasediagram}. As the
localization edge is in the band tail and we are limited to finite-size 
systems, few states are localized. With increasing $\Delta$, the mobility edge
decreases in frequency with respect to the CPA band edge. However, since the band
is broadening, the result is that the critical frequency actually increases 
with disorder, 
and thus there is no vibrational analogue to the electronic MIT. Similar 
behaviour of the mobility
edge with disorder can be seen in the phase diagram of the Anderson model with
off-diagonal disorder \cite{Cain_99}. 

For each degree of disorder, we obtain a new critical spectrum which
is constant for each size. These critical spectra have been
plotted in Fig.\ref{critspecs}, showing that for positive values of q, i.e. the left
hand side of the graph, all the spectra fit onto a master curve. The parabolic
approximation (PA) to Wegner's analytic result \cite{Wegner_89} 
 is one which goes through the 
critical
points $f(\alpha=4)=3$ and $f(\alpha=2)=2$, where the latter corresponds to
the information dimension of the eigenmode. 
This PA has also been plotted on the graph
for comparison. Note that the Wegner result is for the electronic Anderson model, 
yet it still fits well with the vibrational data, indicating a universality
across the two different systems. 
The
large error bars at high $\alpha$ are in the region where $q$ is negative, 
where $f$ and $\alpha$ are strongly dependent on the smallest values of the
measure and where the errors in the eigenmodes themselves are largest.

To conclude, we have investigated the localization phenomenon for vibrational
excitations in disordered structures, using an f.c.c. lattice model with force-constant
disorder for analysis. Using MFA, we have confirmed the existence of only one 
LD transition in the acoustic band, and found the energy at which it occurs for different degrees
of disorder. The eigenmodes at the threshold have been shown to be multifractal
states exhibiting a quantitatively similar distribution function to that of the
critical states in the electron Anderson model.

We are grateful to R.R\"omer for supplying us with MFA code \cite{romercode}, 
and to M.Schreiber for instructive communications.


\begin{thebibliography}{10}
\bibitem{romercode}{Both the German MFA and our own MFA code gave practically
identical results.}
\end{thebibliography}
\end{document}